\documentclass[11pt]{scrartcl}
\pdfoutput=1

\usepackage{graphicx}
\graphicspath{{./}{./Plots/}}
\usepackage{amssymb}
\usepackage{amsmath}
\usepackage{cite}
\usepackage{mathrsfs}
\usepackage{ifpdf}
\usepackage{color}
\usepackage[a4paper, top=1.0in, bottom=1.2in, left=0.8in, right=0.8in]{geometry}

\newcommand{\be}{\begin{equation}}
\newcommand{\ee}{\end{equation}}
\newcommand{\bea}{\begin{eqnarray}}
\newcommand{\eea}{\end{eqnarray}}

\newcommand{\nn}{\nonumber}
\newcommand{\ov}{\overline}
\newcommand{\mc}{\mathcal}

\newcommand{\BRBsmumu}{\BR(B_s \to \mu^+ \mu^-)}
\newcommand{\Bsmumu}{B_s \to \mu^+ \mu^-}
\newcommand{\Bsmm}{B_s \to \mu\mu}
\newcommand{\BR}{{\mathcal B}}

\newcommand{\mcO}{\mc{O}}
\newcommand{\Ell}{\mathscr{L}}

\newcommand{\GeV}{~{\rm GeV}}
\newcommand{\Zbb}{Z\bar{b}b}
\newcommand{\AbFB}{A_{\rm FB}^{0b}}
\newcommand{\dgL}{\delta g_{L}}
\newcommand{\dgR}{\delta g_{R}}
\newcommand{\dgLR}{\delta g_{L,R}}
\newcommand{\DDmu}{{\overset{\leftrightarrow}{D}_\mu}}

\def\sla#1{\setbox0=\hbox{$#1$}\dimen0=\wd0
      \setbox1=\hbox{/} \dimen1=\wd1 \ifdim\dimen0>\dimen1
      \rlap{\hbox to \dimen0{\hfil/\hfil}} #1                        \else
      \rlap{\hbox to \dimen1{\hfil$#1$\hfil}}
      /   \fi}

\begin{document}

\begin{flushright}
LAPTH-010/13\\
CERN-PH-TH/2013-028
\end{flushright}

\medskip

\begin{center}
{\sffamily \bfseries \LARGE \boldmath $\BRBsmumu$ as an electroweak precision test}\\[0.8 cm]
{\large D.~Guadagnoli$^a$ and G.~Isidori$^{b,c}$} \\[0.5 cm]
\small
$^a${\em LAPTh, Universit\'e de Savoie et CNRS, BP110, 
F-74941 Annecy-le-Vieux Cedex, France} \\[0.1cm]
$^b${\em CERN, Theory Division, 1211 Geneva 23, Switzerland} \\[0.1cm]
$^c${\em INFN, Laboratori Nazionali di Frascati, Via E. Fermi 40, 00044 Frascati, Italy}
\end{center}

\abstract{\noindent Using an effective-theory approach, we analyze the impact 
of $\BRBsmumu$ in constraining new-physics models that predict modifications of the 
$Z$-boson couplings to down-type quarks. Under motivated assumptions about the flavor structure 
of the effective theory, we show that the bounds presently derived from $\BRBsmumu$ on the 
effective $Z$-boson couplings are comparable (in the case of minimal flavor violation) or 
significantly more stringent (in the case of generic partial compositeness) with respect to 
those derived from observables at the $Z$ peak.
\noindent 
}

\section{Introduction} \label{sec:intro}

The rare decay $\Bsmumu$ is one of the most clean low-energy probes of physics beyond the Standard 
Model (SM). A first experimental evidence of this rare process has recently been obtained by the 
LHCb collaboration~\cite{Aaij:2012ct}, that reported a $3.5\sigma$ signal. The 
corresponding flavor-averaged time-integrated branching ratio determined by LHCb 
is~\cite{Aaij:2012ct} 
\be
\label{eq:Bsmumu_LHCb}
{\overline \BR}^{\rm exp} = \left(3.2^{+1.5}_{-1.2} \right) \times 10^{-9}~,
\ee
where the error is dominated by the statistical uncertainty and is expected to be improved 
significantly in the near future. At this level of precision there is good agreement with the SM 
prediction, that for the same quantity reads~\cite{Buras:2012ru}
\be
\label{eq:Bsmumu_SM}
{\overline \BR}^{\rm th}_{\rm SM} = \left( 3.54 \pm 0.30 \right) \times 10^{-9}~,
\ee
taking into account the effect of $\Delta\Gamma_s \not=0$ pointed out in Ref.~\cite{Fleischer}. 

The effectiveness of $\Bsmumu$ as a probe of physics beyond the SM is related to a double-suppression 
mechanism at work within the SM. One the one hand, it is a flavor-changing neutral-current (FCNC) process
and, as such, it receives no tree-level contributions. On the other hand, the purely leptonic final state 
and the pseudoscalar nature of the initial state imply a strong helicity suppression and forbid 
photon-mediated amplitudes at the one-loop level. As a result of this double suppression, up to the 
one-loop level $\Bsmumu$ receives contributions only from Yukawa and weak interactions.
 
This process is often advocated as a probe of models with scalar-mediated FCNCs, that are 
naturally predicted in models with an extended Higgs sector. However, it is also an excellent probe 
of the $Z \to b \bar s$ effective coupling 
(see e.g.~Refs.~\cite{Chanowitz:1999jj,Buchalla:2000sk,Haisch:2007ia}).
In this Letter we compare the bounds set on such coupling 
by $\BRBsmumu$ with the deviations from universality on the $Z \to b \bar b $ coupling 
determined from electroweak precision observables. To this purpose, we describe the possible deviations 
on the $Z$-boson couplings to down-type quarks by means of an effective-theory 
approach, and we employ two motivated assumptions about the flavor structure of the theory, namely
minimal flavor violation or generic partial compositeness, to relate flavor-changing and 
flavor-diagonal couplings. 

\section{\boldmath Effective couplings of the $Z$ boson to down-type quarks}

As pointed out in Refs.~\cite{Chanowitz:1999jj,Haisch:2007ia}, there exists a wide class
of models where the only relevant deviations from the SM in $\BRBsmumu$ and $Z \to b \bar b$ can 
be described in terms of modified $Z$-boson couplings at zero momentum transfer, defined by the 
following effective Lagrangian
\be
\Ell^Z_{\rm eff} = \frac{g}{c_W} Z_\mu \ov d^i \gamma^\mu 
\left[ (g^{ij}_{L} +\delta g^{ij}_{L}) P_L + (g^{ij}_{R} +\delta g^{ij}_{R}) P_R \right] d^j~.
\label{eq:LZeff}
\ee
Here $g$ is the $SU(2)_L$ gauge coupling, $c_W=\cos \theta_W$ ($s_W=\sin \theta_W$),
and $g^{ij}_{L,R}$ denote the effective SM couplings. In the following we employ 
state-of-the-art expressions to estimate the SM contributions to $\BRBsmumu$ 
and $Z \to b \bar b$, and use $\Ell^Z_{\rm eff} $ at the tree level only 
to estimate the non-standard effects parameterized by $\delta g^{ij}_{L,R}$.

For later convenience we recall the leading structure of the $g^{ij}_{L,R}$. The tree-level SM 
couplings are 
\be
(g^{ii}_{L})_{\rm tree} = - \frac{1}{2} + \frac{1}{3} s_W^2~, \qquad 
(g^{ii}_{R})_{\rm tree} = \frac{1}{3} s_W^2~, \qquad 
(g^{i\not=j}_{L,R})_{\rm tree} = 0~.
\ee
At the one-loop level the $g^{ii}_{L,R}$ are gauge dependent, but they assume the following 
simple and gauge-independent form in the limit $m_t \gg m_W$ (or $g\to 0$):  
\be
(g^{ij}_{L})_{\rm 1-loop}^{(g=0)} = \frac{m_t^2}{16\pi^2 v^2} V_{ti} ^*V_{tj}~, \qquad 
(g^{ij}_{R})_{\rm 1-loop}^{(g=0)} = 0~,
\label{eq:1LoopSM}
\ee
where $V_{ij}$ denote the elements of the CKM matrix and $v \approx 246$~GeV. 

The new-physics contributions, parameterized by $\delta g^{ij}_{L,R}$, can be related to 
the couplings of a manifestly gauge-invariant Lagrangian,
\be
\Ell^{\rm NP}_{\rm eff} = -\frac{1}{2} \sum_{n,A} \sum_{i,j}
~\frac{ c_{nA}^{ij} }{\Lambda^2} ~ O_{nA}^{ij}~,
\ee
with the following set of dimension-six operators: 
\bea
&& \mcO^{ij}_{1L} = i \left( \ov{Q}^i_L \gamma^\mu Q^j_L \right) H^\dagger \DDmu H~, \qquad 
      \mcO^{ij}_{1R} = i \left( \ov{D}^i_R \gamma^\mu D^j_R \right) H^\dagger \DDmu H~, \nn\\
&&  \mcO^{ij}_{2L} = i \left( \ov{Q}^i_L \tau^a \gamma^\mu Q^j_L \right) H^\dagger \tau^a \DDmu H~.
\label{eq:setops}
\eea
Defining the flavor indices $\{i,j\}$ in the mass-eigenstate basis of down-type quarks we find 
\be 
\delta g^{ij}_{L} = \frac{v^2}{4 \Lambda^2}
\left( c^{ij}_{1L} + \frac{1}{4} c^{ij}_{2L} \right)~, \qquad 
\delta g^{ij}_{R} = \frac{v^2}{4 \Lambda^2} c^{ij}_{1R}~.
\ee

The set of operators in Eq.~(\ref{eq:setops}) is not the complete set of gauge-invariant 
dimension-six operators contributing to $\Bsmumu$ and $Z \to b \bar b$ at the tree level. 
In principle, we can consider also four-fermion (two-quarks/two-leptons) operators, terms of 
the type $J_\nu \times D_\mu F^{\mu\nu}$, or terms of the type 
$H^\dagger J_{\mu \nu} \times F^{\mu\nu}$, where $J_\nu$ and $J_{\mu \nu}$ are quark bilinears, 
and $F^{\mu\nu}$ generically denotes the field-strength tensor of  $U(1)$ or $SU(2)_L$ gauge fields.
However, the effects of these operators cannot be described by means of $\Ell^Z_{\rm eff}$ and 
we lose the natural correlation between these two observables.\footnote{~The four-fermion operators 
do not contribute to $\Ell^Z_{\rm eff}$ at the tree level, hence they have a negligible impact on 
$Z \to b \bar b$ compared to $\Bsmumu$. Conversely, operators with the field-strength tensor generate 
amplitudes suppressed by at least one power of $p/v$, with $p$ the external momentum, that therefore 
have negligible impact on $\Bsmumu$ compared to  $Z \to b \bar b$.}
For this reason in the following we concentrate only on the set of operators in Eq.~(\ref{eq:setops}). 

In order to relate flavor-diagonal and flavor-violating couplings we need to specify the flavor 
structure of the effective theory. We consider two reference frameworks: 1) the hypothesis of Minimal 
Flavor Violation (MFV), as defined in Ref.~\cite{D'Ambrosio:2002ex}; 2) the generic flavor structure 
implied by the hypothesis of Partial Compositeness (PC)~\cite{Kaplan:1991dc}, following the 
effective-theory approach described in Refs.~\cite{Davidson:2007si,KerenZur:2012fr}.

In the MFV framework there is a strict correlation between flavor-diagonal (but non-universal) and 
flavor-violating couplings of the operators listed in Eq.~(\ref{eq:setops}). Restricting to the 
contributions relevant to this correlation, the effective couplings can be decomposed as follows:
\bea
\label{eq:cMFV_L}
(c^{ij}_{nL})^{\rm MFV} &=& a_{nL} \times (Y_u Y_u^\dagger)_{ij} ~\approx~ a_{nL} 
\frac{2 m_t^2}{v^2} V_{ti}^* V_{tj}~, \\
\label{eq:cMFV_R}
(c^{ij}_{1R})^{\rm MFV} &=& a_{1R} \times (Y_d^\dagger Y_u Y_u^\dagger Y_d)_{ij} ~\approx~ a_{1R} 
\frac{4 m_{d_i} m_{d_j} m_t^2 }{ v^4 }  V_{ti}^* V_{tj}~,
\eea
where $a_{nL,R}$ are unknown $O(1)$ couplings and $Y_{u,d}$ are the SM Yukawa couplings. The last 
equalities in Eqs.~(\ref{eq:cMFV_L}), (\ref{eq:cMFV_R}) hold after rotating the Yukawa matrices in 
the mass-eigenstate basis of down-type quarks,  where $Y_u = V^\dagger \lambda_u$ and $Y_d = 
\lambda_d$, with $\lambda_{u,d}$ diagonal  matrices~\cite{D'Ambrosio:2002ex}.

As a result, we can parameterize all the $\delta g^{ij}_{L,R}$ in terms of two flavor-blind parameters, 
$\delta g_{L,R}$, defined by 
\be
(\delta g^{ij}_{L})^{\rm MFV} = \frac{ V_{ti}^* V_{tj}}{|V_{tb}|^2} ~\delta g_L~, \qquad 
(\delta g^{ij}_{R})^{\rm MFV} = \frac{m_{d_i} m_{d_j} }{ m_b^2 } 
\frac{ V_{ti}^* V_{tj} }{|V_{tb}|^2}~\delta g_R~. 
\label{eq:deltaMFV}
\ee
The normalization has been chosen such that
\be
\delta g_{L(R)}^b \equiv \delta g^{33}_{L(R)}   = \delta g_{L(R)}~, 
\label{eq:dgLR}
\ee
in order to identify $\delta g_{L,R}$ with the usual definition of the modified $Z \to b \bar b$ 
couplings~\cite{Hagiwara:1998yc}. As can be seen, in the left-handed sector the flavor structure 
is identical 
to the one of the leading one-loop contribution within the SM, reported in Eq.~(\ref{eq:1LoopSM}). 
In the right-handed sector the structure is different but the effects are expected to be very small 
due to the strong suppression of down-type masses. Indeed the overall normalization implies 
\be
 \delta g_L^{\rm MFV} = \frac{m_t^2 |V_{tb}|^2 }{2 \Lambda^2} 
 \left( a_{1L} + \frac{1}{4} a_{2L} \right)~, \qquad
\delta g_R^{\rm MFV} = \frac{m_b^2 m_t^2 |V_{tb}|^2 }{v^2 \Lambda^2}  a_{1R} ~. \nn
\ee

In the PC framework the correlation between flavor-diagonal and flavor-violating couplings is 
determined up to unknown $O(1)$ parameters, related to the hypothesis of flavor anarchy in the 
composite sector. In this case, following the notation of Ref.~\cite{KerenZur:2012fr}, we expect 
\bea
(c^{ij}_{nL})^{\rm PC} & \sim & \frac{g_\rho^2 \Lambda^2 }{m_\rho^2} \epsilon^q_i \epsilon^q_j 
~\propto ~ |V_{ti} ||V_{tj}|~,
\label{eq:PC1} \\
(c^{ij}_{1R})^{\rm PC} & \sim &  \frac{g_\rho^2 \Lambda^2 }{m_\rho^2}  \epsilon^d_i \epsilon^d_j 
~\propto ~ \frac{m_{d_i} m_{d_j} }{v^2 |V_{ti}||V_{tj}|}~, 
\label{eq:PC2} 
\eea
where the $\epsilon^{q,d}_i$ parameterize the mixing of the SM fermions with the composite sector, and 
$\{m_\rho,g_\rho\}$ are the reference mass and coupling characterizing the composite sector.
On the r.h.s.~of Eqs.~(\ref{eq:PC1}), (\ref{eq:PC2}) we have eliminated the $\epsilon^{q,d}_i$ 
in favor of quark masses and CKM angles by means of the relations~\cite{Davidson:2007si,KerenZur:2012fr}
\be
\frac{ |\epsilon^q_i |}{| \epsilon^q_j | } \sim 
\frac{ |V_{ti} |}{| V_{tj} | } ~, \qquad 
\frac{ |\epsilon^q_i \epsilon^d_i  | }{| \epsilon^q_j  \epsilon^d_j  | } \sim 
\frac{m_{d_i} }{ m_{d_j} }~.
\ee
As can be seen, up to $O(1)$ factors the flavor structure of the left-handed couplings is the same 
as in the MFV framework. On the other hand, the structure is significantly different in the 
right-handed sector, where larger effects are now possible in the flavor-violating case. Ignoring 
$O(1)$ factors, we parameterize the structure of the two couplings in the PC framework as follows:
\be
(\delta g^{ij}_{L})^{\rm PC} = \frac{ |V_{ti}| | V_{tj}| }{|V_{tb}|^2} ~\delta g_L~, \qquad 
(\delta g^{ij}_{R})^{\rm PC} = \frac{m_{d_i} m_{d_j} }{m_b^2}  
\frac{ |V_{tb}|^2}{ |V_{ti} ||V_{tj}|} ~\delta g_R~,
\label{eq:deltaPC}
\ee
where again the normalization has been chosen in order to satisfy Eq.~(\ref{eq:dgLR}). (For recent
studies of the same correlation within specific PC setups, see Ref. \cite{Barbieri}.)
With such choice, the overall normalization implies 
\be
 \delta g_L^{\rm PC} \sim \left(  \frac{ g_\rho \epsilon_3^q   v  }{2 m_\rho  } \right)^2~, \qquad
\delta g_R^{\rm PC}  \sim \frac{1}{2}  \left(  \frac{ m_b  }{\epsilon_3^q  m_\rho  }\right)^2~.
\ee

\section{Analysis and discussion}

The previous considerations can be summarized by stating that, within the two reference frameworks of 
MFV or PC,  possible departures from the SM predictions in the $\Zbb$ couplings 
and in $\BRBsmumu$ can be parameterized in terms of  
the two couplings $\delta g_{L,R}$ defined in Eq.~(\ref{eq:deltaMFV}) or 
Eq.~(\ref{eq:deltaPC}).

Concerning $Z$-peak observables, the $\delta g_{L,R}$  shifts are constrained by $R_b$, $A_b$ and 
$\AbFB$. The state-of-the-art SM calculations for these quantities, to which it is straightforward 
to add the generic shifts in Eq. (\ref{eq:dgLR}), can be implemented following 
Ref.~\cite{Hagiwara:1998yc} (taking also into account the recent SM estimate of $R_b$ in 
Ref.~\cite{Freitas}).
These quantities can then be fitted to the averages of experimental results collected in 
Table~\ref{tab:input}, where we also report the main inputs necessary for their evaluation beyond 
the lowest order.

\begin{table}[t]
\center{
\begin{tabular}{|l|l|}
\hline
& \\
[-0.35cm]
$M_h = 125 \GeV$ \hspace{0.5cm}\hfill \cite{Higgs} & $\Delta \alpha_{\rm had}^{(5)} = 0.02772$ \\
$M_t = 173.2(0.9) \GeV$ \hfill \cite{Tevatron_mt} & $R_b = 0.21629(66)$ \\
$\alpha_s(M_Z) = 0.1184(7)$ \hfill \cite{Bethke_alpha_s} & $A_b = 0.923(20)$ \\
$\alpha^{-1}(M_Z) = 127.937$ \hfill \cite{EWWG_alpha_em} & $\AbFB = 0.0992(16)$ \\
[0.05cm]
\hline
\end{tabular}
}
\caption{Input parameters relevant for  the $Z\to b\bar b$ constraints. Quantities without an explicit
reference are taken from Ref.~\cite{PDG2012}. We do not show the errors for quantities whose uncertainty 
has a negligible impact on our numerical analysis.}
\label{tab:input}
\end{table}

\begin{figure}[t]
\begin{center}
\includegraphics[width=0.48 \textwidth]{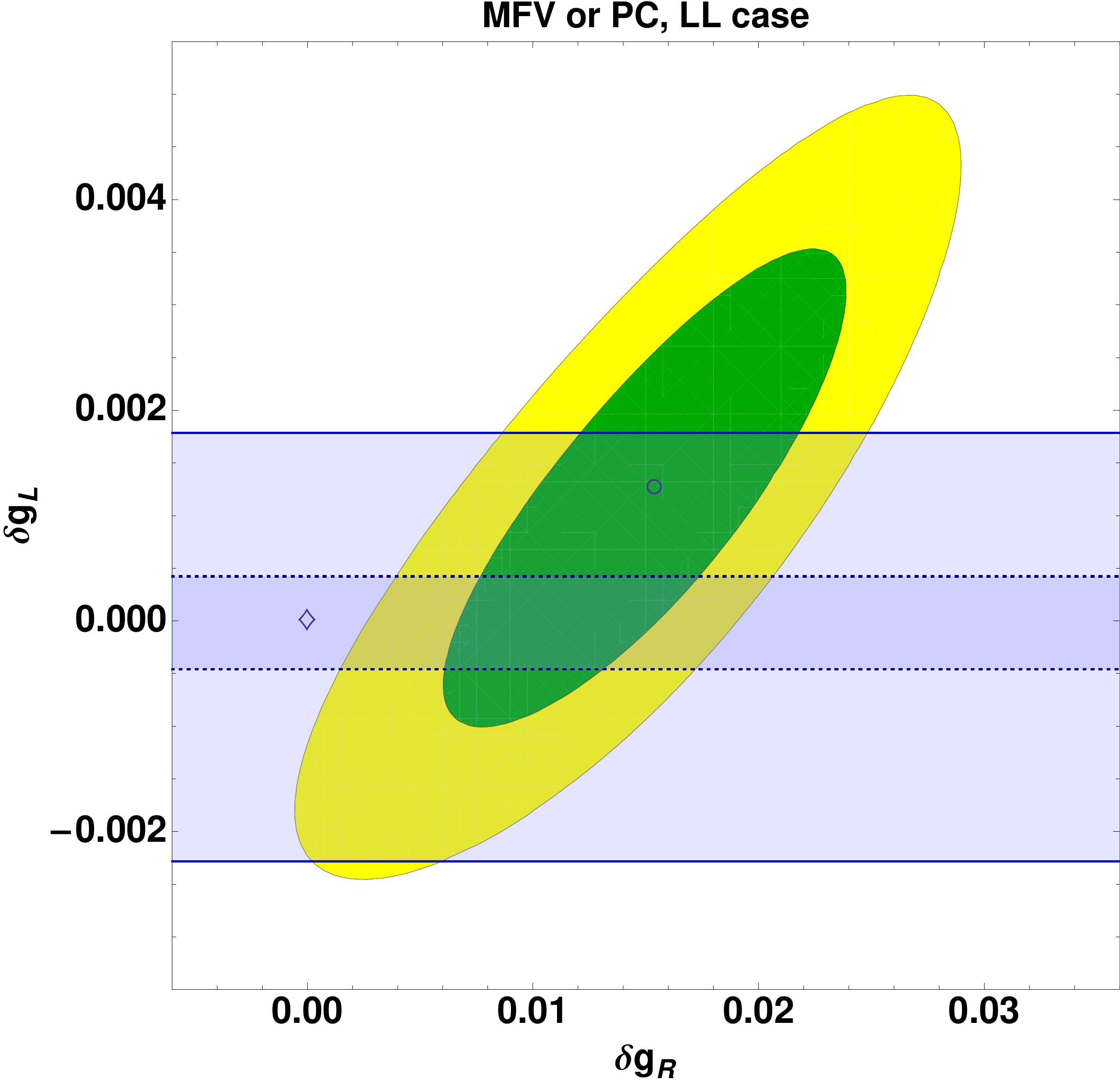}
\includegraphics[width=0.48 \textwidth]{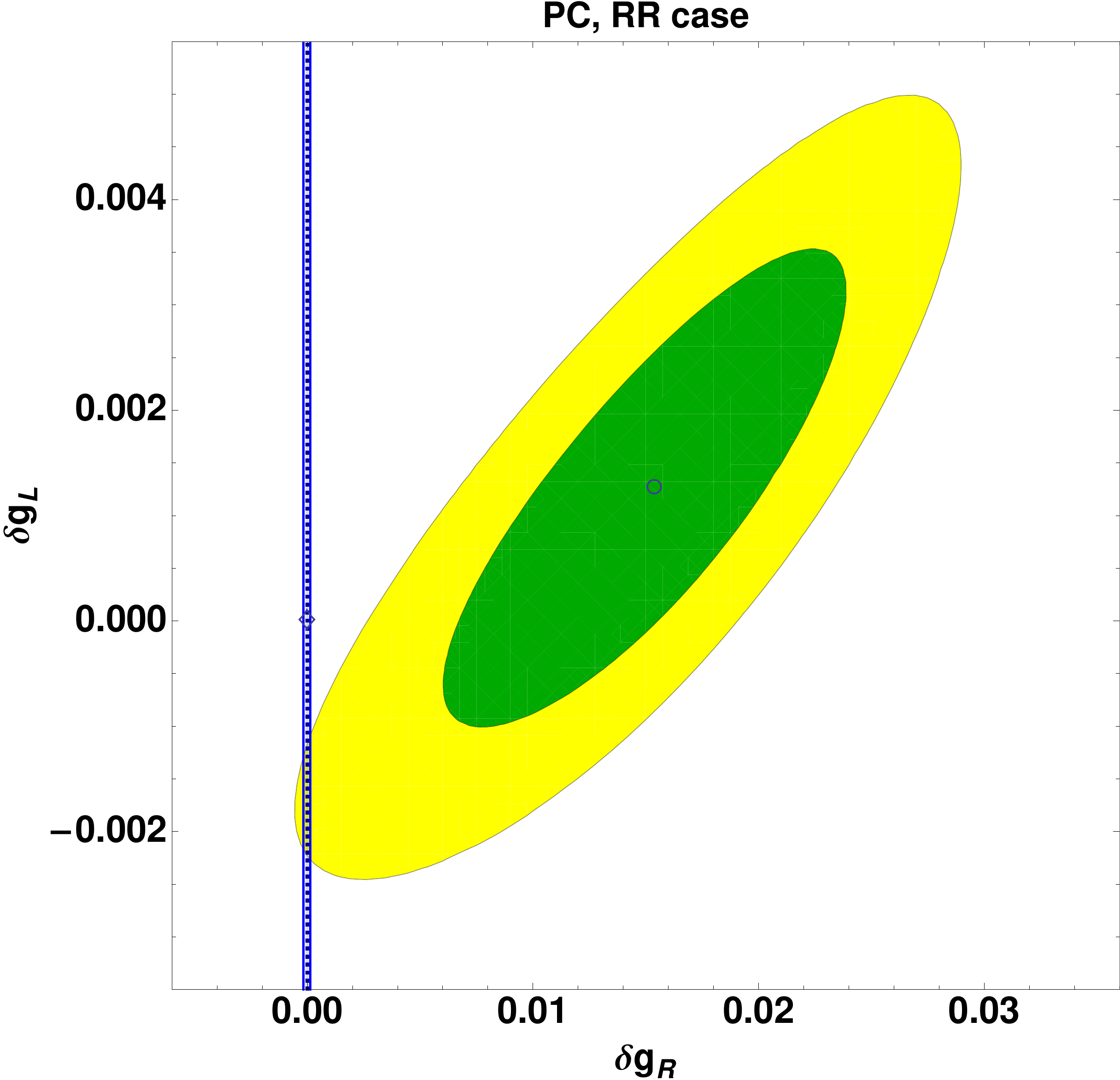}
\end{center}
\caption{Constraints on the couplings $\dgLR$ describing the modified $Z$-boson 
couplings to down-type quarks. The inner and outer ellipses denote respectively the 68\% and 95\% 
CL~regions as obtained from $\Zbb$ observables. The regions delimited 
by solid blue lines denote the 95\% 
CL constraints from $\BRBsmumu$ with the present precision, while 
those comprised between dotted lines are obtained with the $\BRBsmumu$ accuracy expected by 2018
(see text for details).
{\em Left panel:} $\dgL$ constraint from $\BRBsmumu$ under the hypotheses of either MFV or PC.
{\em Right panel:} $\dgR$ constraint from $\BRBsmumu$ under the hypothesis of PC.}
\label{fig:EW_vs_Bsmumu}
\end{figure}

The resulting allowed regions at $68\%$~CL and $95\%$~CL in the $\delta g_{R}$--$\delta g_{L}$ plane 
are shown in~Fig.~\ref{fig:EW_vs_Bsmumu}. As can be noticed, for both $\delta g_{L}$ and $\delta g_{R}$
the fit prefers positive non-zero values, and  the SM point ($\delta g_{R}=\delta g_{L}=0$) is outside 
the $95\%$~CL region. The upper limits for the two parameters are 
\be
\label{eq:dgLR_EW}
|\dgL|_{\Zbb} < 4.5 \times 10^{-3}~, \quad 
|\dgR|_{\Zbb} < 3.0 \times 10^{-2} \qquad [95\%~{\rm CL}]~,
\ee
in  good agreement with the results recently reported in Ref.~\cite{Batell:2012ca}.

Let us now compare these limits with those obtained from the $\BRBsmumu$ measurement within 
the frameworks of MFV or PC. The $\dgLR^{32}$ couplings shift linearly the $Z$-penguin 
contribution to the $\BRBsmumu$ amplitude. These shifts can easily be translated into shifts on 
the short-distance function appearing in the SM formula for the branching ratio (see 
e.g.~Ref.~\cite{Buras:2012ru}). To good accuracy, the effect can simply be described by
\be
\label{eq:BRBsmumu_NP}
\BRBsmumu = \BRBsmumu_{\rm SM} \times \left| 1+ \frac{\sqrt2 \pi^2}{G_F m_W^2 V_{tb}^* V_{ts} } 
\frac{(\dgL^{32} - \dgR^{32})}{Y_{\rm SM}}  \right|^2~,
\ee
where $Y_{\rm SM} \approx 0.957$.\footnote{~A similar expression holds for 
$\BR(\bar B_s \to \mu^+\mu^-)$, with the 
replacement $(\dgL^{32} - \dgR^{32})/V_{tb}^* V_{ts} \to (\dgL^{23} - \dgR^{23})/V_{tb} V_{ts}^*$.
Once $\dgLR^{23,32}$ are expressed in terms of $\dgLR$, the $\BR(\bar B_s \to \mu^+\mu^-)$ and 
$\BR(B_s \to \mu^+\mu^-)$ expressions are identical both in the MFV and in the PC 
parameterization, and can be directly compared with the flavor-averaged branching ratio reported 
by LHCb~\cite{Aaij:2012ct}.
}
Using the 95\%~CL range on the flavor-averaged branching ratio reported by LHCb~\cite{Aaij:2012ct} 
\be
1.1 \times 10^{-9} < {\overline \BR}^{\rm exp} < 6.4 \times 10^{-9}~,
\ee
and the central value of the SM prediction in Eq.~(\ref{eq:Bsmumu_SM}) (at this level of accuracy the 
theoretical error is negligible), one obtains the following bounds on $\dgL$ and $\dgR$:
\bea
&& |\dgL|^{\rm MFV,PC}_{\Bsmumu} < 2.3 \times 10^{-3}~, \nonumber \\
&& |\dgR|^{\rm MFV}_{\Bsmumu} < 1.0 \times 10^{-1}~,
\qquad |\dgR|^{\rm PC}_{\Bsmumu}  < 1.6 \times 10^{-4}~.  
\label{eq:dgLR_Bsmumu_now}
\eea
These bounds have been obtained considering the effects of the two couplings separately (i.e.~barring 
the possibility of cancellations between $\dgL$ and $\dgR$, on which we will comment at the end of 
this section) and ignoring the fine-tuned configuration where the non-standard amplitude is about twice, 
and opposite in sign, compared to the SM one (a possibility that is highly disfavored by the 
 $Z\to b\bar b$ constraints~\cite{Haisch:2007ia}).
 
These bounds are also depicted in Fig.~\ref{fig:EW_vs_Bsmumu} as horizontal or vertical bands delimited 
by solid lines.  From the figure it is evident that, even with its large error, the recent evidence for 
$\BRBsmumu$ provides a constraint on $|\dgL|$ --~under either of the MFV or PC hypotheses~-- 
more stringent than the one obtained from the $Z\to b\bar b$ observables.
Furthermore, the constraint on the $|\dgR|$ coupling within PC is stronger than the one obtained 
from the $Z\to b\bar b$ by more than two orders of magnitude. This circumstance is well represented by 
the right panel of Fig.~\ref{fig:EW_vs_Bsmumu}, where the thickness of the $\BRBsmumu$-allowed band 
(vertical blue `line') is not resolved at the scale of the electroweak-fit ellipse.
This implies that, within anarchic PC models, the $\BRBsmumu$ bound forbids any significant contribution 
to $Z\to b\bar b$ observables able to decrease the existing tension between data and theoretical predictions. 

\medskip 

As far as the bounds on the effective scale of new physics are concerned, in both frameworks 
the constraints derived from the $|\dgL|$ bound in Eq.~(\ref{eq:dgLR_Bsmumu_now})
are largely dominant. They can be summarized as follows: 
\be
\label{eq:Lambda_L_bound}
\Lambda > 2.6~{\rm TeV}\quad  [~{\rm MFV}~(\dgL)~]~, 
\qquad  m_\rho > (g_\rho \epsilon_3^q) \times 2.6~{\rm TeV}\quad [~{\rm PC}~(\dgL)~]~,
\ee
the equality of the numerical coefficient in the two cases being an accident due to the approximate relation
$m_t |V_{tb}|  \approx v/\sqrt2$. 
It is also worth mentioning the $m_\rho$ bound implied by $|\dgR|$ in PC,
\be
\label{eq:m_rho_bound}
m_\rho > \frac{0.23~{\rm TeV}}{\epsilon_3^q} \quad  [~{\rm PC}~(\dgR)~]~,
\ee
that becomes relevant in the limit $\epsilon_3^q \ll1$, in which the bound from $|\dgL|$ gets weaker.

\medskip

While the bounds in Eq.~(\ref{eq:dgLR_Bsmumu_now}) are {\em per se} interesting, the present experimental 
error on $\BRBsmumu$ does not do full justice to the sensitivity of this observable to possible modified 
$Z$-boson couplings. Therefore, we also considered the case of a $\BRBsmumu$ measurement with central 
value as in Eq. (\ref{eq:Bsmumu_SM}) and error of $\pm 0.3 \times 10^{-9}$, that can be considered a 
realistic estimate of the experimental sensitivity on this observable around 2018. This statement takes 
into account the LHCb projections from Ref.~\cite{Bediaga:2012py}, and the fact that CMS will likely 
produce a measurement with similar accuracy. We also assume a still subleading theoretical error, as 
expected by the the steady progress in the lattice determination of the $B_s$ decay constant~\cite{fBs_lat}. 
With these assumptions on the projected total error on $\BRBsmumu$, the 95\%~CL bounds on $\dgLR$ become 
\bea
&& |\dgL|^{\rm MFV,PC}_{[\sigma(\Bsmm)=3\times 10^{-10}]} < 4.6 \times 10^{-4}~, \nonumber \\
&& |\dgR|^{\rm MFV}_{[\sigma(\Bsmm)=3\times 10^{-10}]} < 2.0 \times 10^{-2}~, 
\quad |\dgR|^{\rm PC}_{[\sigma(\Bsmm)=3\times 10^{-10}]} < 3.3 \times 10^{-5}~,  \qquad 
\label{eq:dgLR_Bsmumu_2018}
\eea
and the bounds in Eqs. (\ref{eq:Lambda_L_bound}) and (\ref{eq:m_rho_bound}) improve by
a factor of about two.
The comparison between Eq.~(\ref{eq:dgLR_Bsmumu_2018}) and  Eq.~(\ref{eq:dgLR_EW}) illustrates the 
potential of uncovering even tiny new-physics deviations in the $Z$-boson couplings to down-type quarks 
via $\BRBsmumu$. Note that, in the pessimistic case where no deviations from the SM prediction are 
observed in $\BRBsmumu$, even the bound on $\dgR$ within MFV will become more stringent compared to 
the one obtained from the $Z\to b\bar b$ observables.

Besides improvements in the $\BRBsmumu$ measurement, a further avenue towards reducing the error 
on the $Z\to b\bar s$ effective coupling is, in principle, that of combining the constraints from 
other $b \to s$ decays, most notably $B\to K^*\mu^+\mu^-$ and $B\to K \mu^+ \mu^-$ (recent attempts in 
this direction can be found in Ref.~\cite{Bobeth:2011nj}; see also Ref.~\cite{Hiller} for other related
studies).
However, the extraction of information about the $Z \to b\bar s$ effective coupling from these decays is 
not as pristine as in the $\BRBsmumu$ case. In fact, on the one side, and at variance with $\Bsmumu$, 
these processes receive, already within the SM, substantial contributions from amplitudes other than 
the $Z$-penguin. In addition, the definition of observables related to these processes comes 
with inevitable theoretical assumptions, related to the dependence on additional hadronic form factors.

Finally, as anticipated, the bounds in Eq.~(\ref{eq:dgLR_Bsmumu_now}) and Eq.~(\ref{eq:dgLR_Bsmumu_2018}) 
do not take into account the possibility of cancellations in the case where {\em both} $\dgL$ and $\dgR$ 
are switched on simultaneously.
In practice, admitting such possibility does not lead to any significant changes in the plots of 
Fig.~\ref{fig:EW_vs_Bsmumu}. As expected from the hierarchical nature of the bounds in Eqs. 
(\ref{eq:dgLR_Bsmumu_now}) or (\ref{eq:dgLR_Bsmumu_2018}), the allowed region in the case of 
simultaneously non-zero $\dgL$ and $\dgR$ is dominated by the region allowed by the strongest 
constraint, namely $\dgL$ in the case of MFV and $\dgR$ in the case of PC.

\bigskip

\section{Conclusions}

The long-standing discrepancy between experimental data and SM predictions for the $Z\to b \bar b$ 
observables ($\AbFB$ and, {to a lesser extent, also $R_b$) has often been advocated as a possible hint 
of physics beyond the SM. If this is the case, under reasonable assumptions about the flavor structure 
of the new-physics model, sizable non-standard contributions should also be expected in $\Bsmumu$.

A first attempt to relate flavor-changing and flavor-diagonal constraints on the $Z$-boson couplings, 
under the assumption that they provide the dominant new-physics contribution to both $\BRBsmumu$ and 
$Z \to b \bar b $, was made in Ref.~\cite{Haisch:2007ia}. At that time, the information from 
$Z \to b \bar b $ observables was used to derive possible upper bounds on $\BRBsmumu$ and other FCNC 
processes. The situation is now reversed: the experimental precision reached on $\BRBsmumu$ is such that 
this observable sets the dominant constraints on possible modified $Z$-boson couplings.

In MFV models, where sizable deviations are expected only in the left-handed couplings of the $Z$ boson, 
the bound presently derived from $\BRBsmumu$  is only slightly more stringent with respect to the one 
derived from $Z \to b \bar b$. However, the situation is likely to improve soon with the foreseen 
 experimental progress on $\BRBsmumu$, see Fig.~\ref{fig:EW_vs_Bsmumu} left. 
In generic models with partial compositeness, $\BRBsmumu$ sets a constraint on possible modifications 
of the right-handed coupling considerably more stringent than $Z \to b \bar b$, see Fig.~\ref{fig:EW_vs_Bsmumu} right.  
This constraint forbids any significant contribution to  $Z\to b\bar b$ observables able to decrease 
the existing tension between data and theoretical predictions.

More generally, our results illustrate how a measurement of $\BRBsmumu$ with the expected
accuracy of order 10\% is able to unveil even tiny new-physics deviations in the $Z$-boson 
couplings to down-type quarks.

\section*{Acknowledgments}
We thank Ayres Freitas, Stefania Gori, Matteo Palutan, Ennio Salvioni and Barbara Sciascia 
for useful discussions and feedback.
The work of DG was partially supported by a PEPS PTI Grant. 
The work of GI was supported by the EU ERC Advanced Grant FLAVOUR (267104), and by MIUR under 
project 2010YJ2NYW.

\end{document}